\documentclass[twoside]{article}
\usepackage[preprint]{aistats2027}
\usepackage{graphicx,algorithm,algorithmic,booktabs,wrapfig,varwidth}
\usepackage{placeins}
\usepackage{amsmath,amssymb,amsthm,natbib,hyperref,url}
\hypersetup{
  hidelinks,
  pdftitle={TTHE: Test-Time Harness Evolution},
  pdfauthor={Jun Nie; Yonggang Zhang; Jun Song; Qianshu Cai; Dahai Yu; Yike Guo; Xinmei Tian; Bo Han}
}
\begin{document}
\runningtitle{TTHE: Test-Time Harness Evolution}
\runningauthor{TTHE: Test-Time Harness Evolution}
\twocolumn[
\aistatstitle{TTHE: Test-Time Harness Evolution}
\aistatsauthor{
  Jun Nie\textsuperscript{1,2},
  Yonggang Zhang\textsuperscript{3,*},
  Jun Song\textsuperscript{1},
  Qianshu Cai\textsuperscript{2},
  Dahai Yu\textsuperscript{4},\\[2pt]
  \bfseries Yike Guo\textsuperscript{3},
  Xinmei Tian\textsuperscript{2},
  Bo Han\textsuperscript{1,*}
}
\aistatsaddress{
  \textsuperscript{1}Hong Kong Baptist University\quad
  \textsuperscript{2}University of Science and Technology of China\\[2pt]
  \textsuperscript{3}Hong Kong Generative AI Research \& Development Center,\\
  The Hong Kong University of Science and Technology\\[2pt]
  \textsuperscript{4}TCL Corporate Research (HK) Co., Ltd
}]
\begingroup
\makeatletter
\renewcommand{\@makefntext}[1]{\noindent\raggedright\@makefnmark#1}
\makeatother
\renewcommand{\thefootnote}{\fnsymbol{footnote}}
\footnotetext[1]{Corresponding authors.\\
  Code: \url{https://github.com/junnie00/TTHE}}
\endgroup
\begin{abstract}
The behavior of an LLM agent is determined not only by the underlying model, but also by its \emph{harness}: the executable program that constructs context, invokes tools, verifies intermediate results, and recovers from failures. Existing approaches typically optimize harnesses on training or development data and freeze the resulting programs at test time, limiting adaptation to shifts in task distributions, failure modes, and tool interactions. We ask whether the harness can instead be optimized during evaluation using unlabeled execution traces. We introduce \emph{Test-Time Harness Evolution} (TTHE), which treats the executable harness as the state of test-time adaptation. On each unlabeled test batch, TTHE iteratively evolves a population of candidate harnesses through agentic proposers that reason over execution traces and explore diverse edit directions. A judge selects a candidate using execution traces and execution-derived proxy signals. The selected harness determines the current batch's outputs and initializes adaptation on the next. Crucially, the solver, proposers, and judge operate through different harnesses around the same frozen LLM, without weight updates, access to final-answer labels, or a separately trained adaptation model. Across six execution-grounded benchmarks spanning text-to-SQL, competitive programming, software engineering, data-science coding, terminal-based tasks, and agentic tool use, TTHE improves fixed ReAct-style baseline harnesses through persistent, inspectable program adaptation. These results recast test-time adaptation for LLM agents as evolution over executable control programs and identify execution-derived proxy reliability as a central challenge for robust unsupervised agent improvement.
\end{abstract}

\section{Introduction}

Modern LLM agents are increasingly built as executable systems rather than single-call predictors.
Their behavior depends not only on the model, but also on the executable \emph{harness} around
it---the control program that constructs context, invokes tools, verifies intermediate results,
and implements recovery strategies. Following prior work on model and agent
harnesses~\citep{lee2026metaharness,zhang2026selfharness}, we focus on this program as the object
of adaptation. Its context construction and control flow determine how the model's capabilities
are used. A common development workflow, however, is to inspect failures, adjust prompts and
control flow, and tune tool-use logic before evaluating the resulting harness as a fixed artifact.
Its operating procedure then remains unchanged even when the target workload exposes different
requirements or failure modes.

This convention treats evaluation as a scoring stage, even though execution itself produces rich
operational evidence. Every attempted task leaves an execution trace: model calls, tool actions,
intermediate outputs, database responses, test results, runtime errors, and recovery decisions.
Such traces can reveal not only whether an execution satisfies available checks, but also how
its operating procedure might be improved. A text-to-SQL agent may retrieve the correct table
but apply an inconsistent value normalization; a coding agent may generate plausible patches
without verifying them before submission; a data-science agent may compute an intermediate
result but omit it from the final report. This is precisely the kind of evidence a human harness engineer would inspect when
revising the scaffold. Crucially, this evidence is available during evaluation without
access to gold labels.

Using this evidence to revise the harness, however, is different from using it to
refine individual responses. Self-debugging and reflection use feedback to revise
candidate answers or retain textual experience for later
attempts~\citep{chen2024selfdebug,shinn2023reflexion}. Automated prompt and workflow
search targets the surrounding task-solving procedure, but typically optimizes it
against a training or development objective before evaluating the resulting workflow
as a fixed artifact~\citep{khattab2024dspy,hu2025adas,zhang2025aflow}. More recent
work revises agent source code, harnesses, and configurations using execution feedback
or task-level evaluation
signals~\citep{moss2026,lee2026metaharness,zhang2026selfharness,he2026evotest}.
Building on these lines of work, we ask:
\emph{can an LLM agent adapt its harness at test time using its own execution traces,
without final correctness feedback?}

We propose \emph{Test-Time Harness Evolution} (TTHE), which treats the executable harness as
the state of test-time adaptation. Rather than searching for a harness on labeled development
instances and freezing it for evaluation, TTHE moves harness optimization into the evaluation
process itself. On each unlabeled test batch, TTHE executes a population of candidate harnesses
and iteratively refines them through trace-guided proposers that explore diverse edit directions.
A judge selects one candidate using the traces and execution-derived proxy signals; the selected
harness determines the current batch's outputs and initializes adaptation on the next.
Final correctness feedback is used only for post-selection evaluation. The solver, proposers, and judge use the same frozen LLM in
different roles, so adaptation changes the executable program rather than model weights and
requires no separately trained adaptation model.

This evaluation-time formulation offers two practical advantages. First, it adapts at the point
where the relevant evidence appears: after the agent encounters the schemas, repositories,
tests, tools, and runtime constraints of the target workload. It can therefore use task-specific
execution evidence without requiring an additional labeled development set tailored to that
workload. Second, updating an executable program requires no gradient access and permits
explicit changes to context construction, multi-sample generation, tool-mediated grounding,
deterministic contract checks, and conditional recovery branches. These changes alter the
procedure used to solve tasks rather than only revising an individual response.

A central challenge, however, is that execution evidence is not a correctness oracle. A query
can execute successfully while answering the wrong question; code can pass visible tests while
failing hidden ones. TTHE therefore presents proxy signals together with candidate code and
the underlying execution traces, allowing proposers and the judge to interpret these signals
in the context of how an outcome was produced rather than relying on a single surrogate score.
Because an edit intended to repair a failure can also disrupt previously correct behavior,
we complement this design with a conditional analysis of a class of trace-triggered harness
edits, characterizing when corrective gains outweigh unintended regressions.

We evaluate TTHE on six execution-grounded benchmarks, where it
improves fixed ReAct-style baselines and yields inspectable grounding, verification, and recovery
policies. On the evaluated BIRD dataset, TTHE achieves $48.7\%$ mean execution
accuracy over three runs, compared with $12.7\%$ for the fixed baseline. A separate ten-run
execution-result consistency-voting control achieves $16.0\%$. Harnesses selected by DeepSeek for each BIRD batch
also improve MiMo\,V2.5 and GPT-5.5 over their respective fixed-harness baselines on the
corresponding batches, without further harness evolution. These results provide evidence of
cross-model transfer of the evolved task-solving procedures. Search diagnostics reveal
non-monotonic effects of search budget and distinguish candidate-availability limitations from
selection errors. Together, these findings support executable harness adaptation as a practical
path to test-time self-evolution in LLM agents.

\noindent\textbf{Contributions.}
\begin{itemize}
\item We introduce \textbf{Test-Time Harness Evolution (TTHE)}, a framework that
recasts test-time adaptation as evolution over executable control programs.
TTHE enables an LLM agent to transform its own execution experience into
harness updates, without modifying model weights or accessing final
correctness feedback.
\item We demonstrate gains on evaluated tasks across six execution-grounded benchmarks,
compare TTHE with fixed-harness repeated sampling on BIRD, and show that batch-specific
evolved harnesses can benefit other execution models on the corresponding tasks.
\item We present inspectable harness examples, provide a conditional analysis of corrective
gains and unintended regressions for a class of trace-triggered edits, and use search
diagnostics to distinguish candidate availability from selection errors.
\end{itemize}

\section{Problem Setting}
\label{sec:setting}

Let $M$ be a frozen LLM and $\mathcal H$ the space of executable harnesses.
A harness $H\in\mathcal H$ orchestrates calls to $M$ and interactions with an
allowed task environment $E$. Executing $H$ on input $x$ produces an outcome
$o_H(x)$ and a trace $\tau_H(x)$, with execution randomness suppressed in the
notation. Outcomes include submitted artifacts and final environment states
relevant to evaluation, not only the agent's final message.

\paragraph{Current-batch adaptation.}
At step $t$, the agent receives an unlabeled batch
$X_t=(x_{t,1},\ldots,x_{t,B_t})$ and an incoming harness $H_t$.
Starting from $H_t$, the adaptation procedure interacts with $E$ to construct,
execute, and select a candidate harness:
\begin{equation}
 (\widetilde H_t,\widehat O_t)
 =\operatorname{Adapt}(M,H_t,X_t,E).
 \label{eq:adaptation}
\end{equation}
Here $\widehat O_t=(\widehat o_{t,1},\ldots,\widehat o_{t,B_t})$ contains the
selected candidate's cached outcomes on $X_t$. After commitment, these outcomes
are evaluated against the hidden targets $Y_t=(y_{t,1},\ldots,y_{t,B_t})$,
without re-execution or reselection:
\begin{equation}
 \widehat J_t=\frac{1}{B_t}\sum_{i=1}^{B_t}
 \operatorname{score}(\widehat o_{t,i},y_{t,i}).
 \label{eq:batchscore}
\end{equation}
The task-specific function $\operatorname{score}$ takes values in $[0,1]$.
For binary tasks, it is $\mathbb I[\widehat o_{t,i}\equiv y_{t,i}]$, where
$\equiv$ denotes task correctness, such as SQL result-set equality or passing
all hidden tests; graded tasks use their evaluation rubric.
TTHE sets $H_{t+1}=\widetilde H_t$ to initialize adaptation on the next batch.
Evaluation is \emph{transductive}: the current batch supplies unlabeled
adaptation evidence and is scored after adaptation, rather than using the
incoming harness $H_t$ before adaptation.

\paragraph{Information available to adaptation.}
The adaptation loop may access task inputs, candidate source code, execution
traces, and permitted environment responses. Public examples or repository
tests provide partial checks when exposed by the task. Final-answer labels,
reference solutions, hidden-test outcomes, and final evaluation scores remain
inaccessible to all adaptation components and their traces; evaluation
feedback is not passed to subsequent batches. No calibrated scalar
correctness reward is assumed.

Let $\mathcal F_t\subseteq\mathcal H$ be the final candidate set eligible for
selection, and let $\mathcal T_t(H)$ collect the execution traces and
execution-derived signals for candidate $H$ on $X_t$. A label-free judge
$\mathcal J$ selects a single harness $\widetilde H_t\in\mathcal F_t$:
\begin{equation}
 \widetilde H_t=\mathcal J\!\left(
 X_t,\{(H,\mathcal T_t(H)):H\in\mathcal F_t\}\right).
 \label{eq:select}
\end{equation}
The judge may acquire additional evidence through permitted probes in $E$,
subject to the same information constraints.

\section{Method}
\label{sec:method}
\subsection{Overview}
TTHE adapts an agent's executable harness using evidence generated during test-time execution.
Two questions guide its design: \textbf{(Q1)}~What evidence can support adaptation without final
correctness feedback? \textbf{(Q2)}~How can this evidence guide program revision and candidate
selection? We address Q1 with execution traces and task-dependent partial checks, and Q2 with
iterative evolution of multiple harness branches followed by batch-level selection.
Starting from $H_t$, the loop evolves $G$ branches over $R$ rounds on the current batch $X_t$.
Figure~\ref{fig:overview} illustrates one batch; Algorithm~\ref{alg:tthe} in
Appendix~\ref{app:algorithm} gives the full procedure.

\begin{figure*}[t]
\centering
\includegraphics[width=0.9\textwidth]{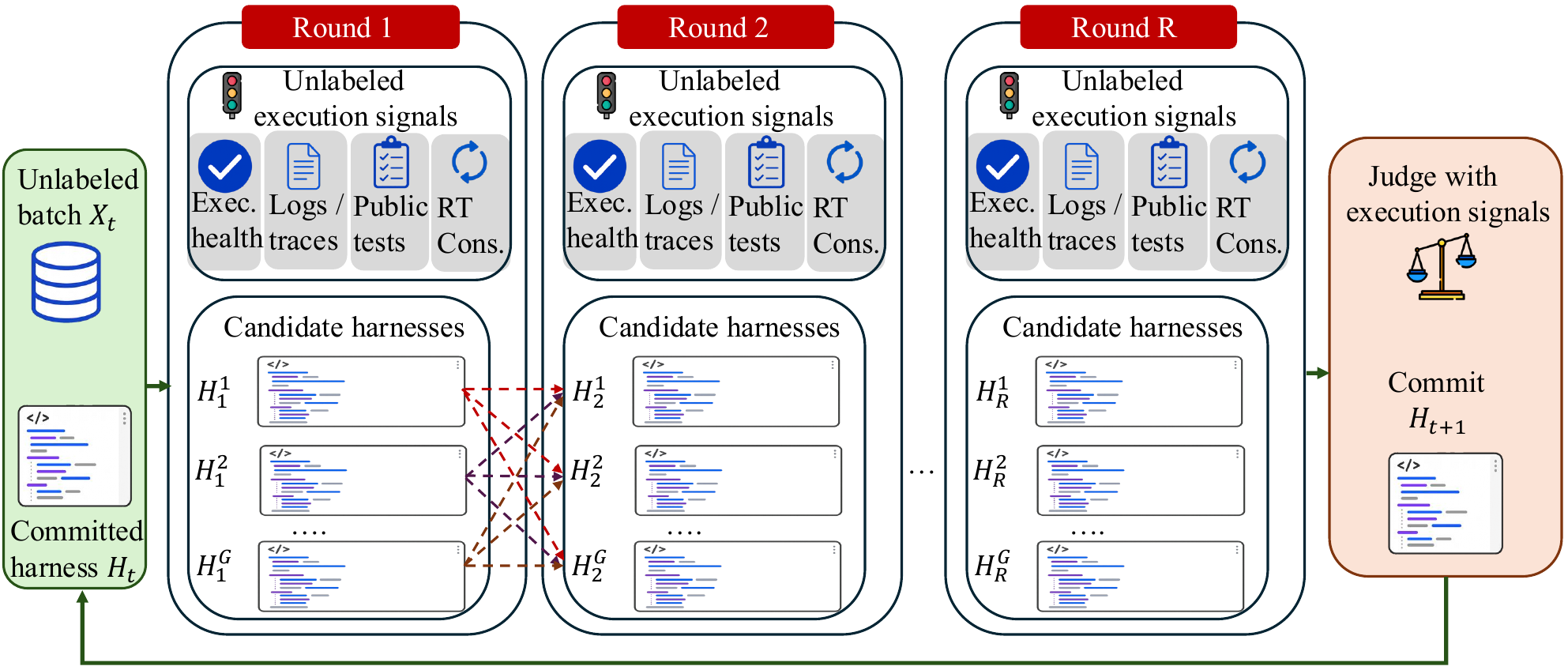}
\caption{Overview of TTHE.
On each unlabeled batch $X_t$, proposers evolve $G$ harness branches from $H_t$
over $R$ rounds, exploring diverse edits guided by execution traces.
A judge selects one final-round candidate using traces and execution-derived proxy signals.
The selected harness determines the current batch's outputs and initializes adaptation
on the next batch as $H_{t+1}$. Final correctness feedback is excluded from the loop.}
\label{fig:overview}
\end{figure*}

\subsection{Harness as adaptation state}
The adaptation state is the task harness's source, including its prompts and executable control
logic. TTHE revises how the harness constructs context, invokes tools, validates intermediate
results, and recovers from failures, while keeping the model and the outer proposal-and-selection
procedure fixed. The baseline $H_0$ is a ReAct-style scaffold~\citep{yao2023react}, fixed before
evolution begins.

A code edit can change a prompt, introduce a database probe before generation, add a deterministic
output check, or create a conditional repair branch. It can also implement multiple-sample generation
and aggregation~\citep{wang2023selfconsistency} or execution-based self-debugging~\citep{chen2024selfdebug}.
The search is therefore not confined to choosing among predeclared prompt templates or workflow settings.
The distinction from response-level refinement is the object of the update: a revised query is an output,
whereas a new rule for constructing and checking queries changes the procedure applied to the batch.

\subsection{Execution Evidence}
\label{subsec:signal}
Executing a candidate records its model prompts and completions, tool calls and responses, stdout/stderr,
intermediate artifacts, runtime states, errors, and probe results; long fields are truncated. A trace
exposes how a harness operates: which tools it calls, whether intermediate artifacts are consumed,
when validation or recovery branches fire, and how the final output relates to the preceding computation.
These details provide context for diagnosing a failure and proposing changes to the operating procedure.
For an input $x$ and outcome $o$, TTHE also exposes task-dependent proxy signals:
\begin{itemize}
\item \textbf{Execution health} $s\in\{0,1\}$ records whether the output executes and satisfies available
format or shape checks, such as returning a scalar for a counting query.
\item \textbf{Round-trip consistency} $b\in[0,1]$ compares the input to a description reconstructed from
the output: $\widetilde x=\operatorname{RT}(o)$ and $b=\operatorname{match}(x,\widetilde x)$. A mismatch
provides evidence of possible semantic drift without requiring a reference answer.
\item \textbf{Public-test pass rate} $p\in[0,1]$ records the fraction of exposed examples or repository
tests passed, where such tests are available.
\end{itemize}
These signals provide complementary but incomplete evidence: successful execution does not establish
semantic correctness, and different candidates may agree on an incorrect result. TTHE therefore
presents the signals together with candidate code and the underlying execution traces, rather than
aggregating them into a single scalar objective. Proposers use this evidence to diagnose weaknesses
and propose program changes, while the judge uses it to compare the resulting candidates.

\subsection{Evolution and Selection}
\label{subsec:evolve}
Let $H_r^g$ denote branch $g$ after round $r$, with $H_0^g=H_t$ for all $g$.
The loop uses three operators under the information constraints in Section~\ref{sec:setting}:
\begin{description}
\item[\textsc{Observe}$(H,X_t)$] executes harness $H$ on every input in $X_t$ and records its
execution traces, proxy signals, and cached outcomes.
\item[\textsc{Propose}$(\mathcal A,\mathcal T)$] receives the active branches $\mathcal A$ and their
execution records $\mathcal T$. One coding-agent proposer per branch revises its own parent,
using the active branches' code and traces for context, and returns one child.
\item[\textsc{Judge}$(\mathcal F,\mathcal T,X_t)$] inspects the eligible candidates' code and signals. During selection, it may acquire additional evidence through probes or candidate
re-execution, then commits one harness from $\mathcal F$.
\end{description}
At the start of a batch, the incoming harness is observed once, and its cached execution record is
shared across the initial branches. In each round, valid children replace their parents and are
executed on $X_t$; invalid or unloadable children leave their parents and cached records unchanged. Two design choices organize these updates. \emph{Lineage}: a fixed parent--child sequence allows
successive edits to build on the branch's grounding, verification, and recovery routines.
\emph{Diversity}: conservative-repair, exploratory, and adversarial proposer roles encourage
different edit directions. Branches can use one another's execution evidence while retaining their
own lineages. After $R$ rounds, the judge selects one harness from
$\mathcal F_t=\{H_R^g\}_{g=1}^{G}$ based on the candidates' code,
execution traces, and proxy signals. The selected harness's cached outcomes
determine the current batch's score in Eq.~\ref{eq:batchscore}, and the
harness initializes adaptation on the next batch.

The solver, proposers, and judge share the same frozen LLM. The solver runs within
the evolving task harness, while the proposers and judge use a Claude Code scaffold
for file access, code editing, execution, and probing, with all model calls routed
to the same backbone.

\section{Theoretical Analysis}
\label{sec:theory}
We analyze a class of trace-triggered harness edits on the current batch, rather than the outer
proposal-and-selection process. As conditional analyses of test-time training relate an auxiliary update
to the main-task objective~\citep{sun2020ttt}, this analysis relates corrective edits to regressions.

\subsection{Setup and Assumptions}
\label{subsec:editmodel}
Fix a nonempty batch $X=\{x_i\}_{i=1}^{B}$, its hidden evaluation targets, and a realized label-free
construction history $z$, including the random choices used to construct $H,H'$ and a trigger rule.
Hold these programs, the trigger rule, and any inherited state fixed. Draw $I$ uniformly from
$\{1,\ldots,B\}$ and choose a fixed coupling $\kappa$ of fresh execution records $(O,\tau)$ and
$(O',\tau')$ on $x_I$. Both joint output--trace marginals must match the corresponding fixed program's
execution law in the same prescribed initial environment. Outcomes include submitted artifacts and
relevant final states. With $y=y_I$, define binary correctness and its gain by
\begin{equation}
\begin{gathered}
\begin{aligned}
C &= \mathbb{I}[O \equiv y],
& C' &= \mathbb{I}[O' \equiv y],\\
e &= \Pr(C=0),
& e' &= \Pr(C'=0),
\end{aligned}\\[2pt]
\Delta_X(H,H') = e-e'.
\end{gathered}
\label{eq:editrisk}
\end{equation}
Probabilities average over $I$ and fresh execution randomness conditional on the fixed history;
dependence on $z$ and $\kappa$ is suppressed. Gold defines correctness for analysis only.

Let $D=\mathsf d(x_I,\tau)\in\{0,1\}$, where the fixed, label-free rule $\mathsf d$ identifies executions
targeted by the edit. This execution-time trigger is distinct from the historical evidence that motivated
the source-code revision. Define
\begin{equation}
 \varepsilon=\Pr(D=0,\ O'\ne O).
 \label{eq:locality}
\end{equation}
Exact locality means $\varepsilon=0$: untriggered outcomes are preserved under $\kappa$.
The value of $\varepsilon$ depends on the execution coupling, not only on the source-code change.
For example, a prefix-preserving repair can have the form
\[
 O'=\begin{cases}
 O, & D=0,\\
 \operatorname{Repair}(x_I,O,\tau;\xi), & D=1,
 \end{cases}
\]
where the child retains the parent execution before the repair decision and $\xi$ supplies fresh
repair randomness. Sharing that execution gives exact locality if the check leaves untriggered
outcomes unchanged. This illustrates an editable program structure, not a constraint imposed by TTHE;
a global prompt or grounding edit need not have this form. Locality is a behavioral condition, not a
limit on edited source lines. Appendix~\ref{app:theory} specifies the coupling and its assumptions.

Separate how often a trigger fires from how informative it is and what the edit does:
\begin{equation}
\begin{aligned}
 m&=\Pr(D=1),\\
 \pi&=\Pr(C=0\mid D=1),\\
 r&=\Pr(C'=1\mid C=0,D=1),\\
 u&=\Pr(C'=0\mid C=1,D=1).
\end{aligned}
\label{eq:editrates}
\end{equation}
Here $m$ is trigger mass, $\pi$ is its precision as an error detector, $r$ is repair success, and $u$ is
regression after falsely triggering on a correct outcome. Rates on zero-probability conditioning events
are set to zero. The rates $r,u$ and the locality term $\varepsilon$ refer to the specified coupling;
these are analysis quantities, not correctness scores accessible to the loop.

\newtheorem{ttheproposition}{Proposition}
\newtheorem{tthetheorem}{Theorem}

\subsection{Conditions for Improvement}
\begin{ttheproposition}[Corrective gains and unintended regressions]
\label{prop:balance}
Under the coupled execution model,
\begin{equation}
 \left|\Delta_X(H,H')-m[\pi r-(1-\pi)u]\right|\leq\varepsilon.
 \label{eq:editbound}
\end{equation}
Hence $m\pi r>m(1-\pi)u+\varepsilon$ suffices for strict improvement. With exact locality,
$\Delta_X(H,H')=m[\pi r-(1-\pi)u]$, so $m\pi r>m(1-\pi)u$ is necessary and sufficient.
\end{ttheproposition}

The balance separates successful repairs, false-trigger regressions, and off-trigger changes. The last
term can help or harm; $\varepsilon$ bounds its magnitude rather than measuring an unavoidable loss.
The sufficient condition requires repairs to outweigh false-trigger regressions and worst-case
off-trigger harm. Trigger precision alone does not establish repair success.

\subsection{Preservation and Feedback Limits}
A trigger is \emph{sound for rejection} when $D=1$ implies $C=0$ almost surely. Violating an authoritative
output contract, or failing a public test that is a necessary condition for task correctness, can provide
such evidence without identifying the correct answer. An empty SQL result, a recovered intermediate
exception, or low semantic consistency does not in general have this property.

\begin{tthetheorem}[Preservation and the limit of failure feedback]
\label{thm:soundrepair}
If the trigger is sound for rejection and the edit is exactly local, then
\begin{equation}
 \Delta_X(H,H')=\Pr(D=1,C'=1)=mr\geq0.
 \label{eq:soundsafety}
\end{equation}
No independence or positive lower bound on repair success is required.

If a label-free child check $D'$ is also sound for rejection and
$v=\Pr(D=1,D'=0)$ is the mass of cleared failure flags, then
\begin{equation}
 0\leq\Delta_X(H,H')\leq v.
 \label{eq:clearancebound}
\end{equation}
Both endpoints are attainable with the same distribution of $(D,D')$ under sound but incomplete checks.
Thus failure-flag clearance alone cannot certify a strictly positive correctness gain.
\end{tthetheorem}

\begin{figure*}[t]
\centering
\includegraphics[width=0.9\textwidth]{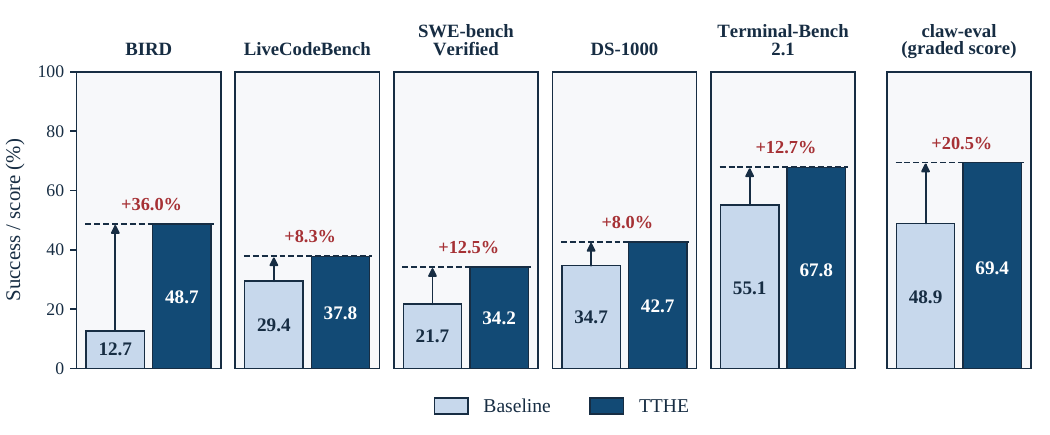}
\vspace{-0.3cm}
\caption{\textbf{TTHE versus fixed baseline harnesses across six benchmarks.}
Bars show mean task success rates (\%) for the first five benchmarks and
mean graded scores for claw-eval. Red annotations indicate
absolute gains over the baseline. Individual runs and mean $\pm$ standard
deviation are provided in Appendix~\ref{app:repeated-results}.}
\label{fig:main}
\end{figure*}

The preservation result confines modifications to outcomes already known to be wrong; strict progress
still requires successful repair. With imperfect triggers, Proposition~\ref{prop:balance} gives
$e'\leq e+m(1-\pi)+\varepsilon$ without a repair-quality assumption. Complete proofs, equivalent
parameterizations, and sharpness constructions appear in Appendix~\ref{app:theory}.

\section{Experiments}
\label{sec:experiments}
\subsection{Experimental Setup}
We evaluate TTHE on six execution-grounded benchmarks: BIRD~\citep{li2023bird},
LiveCodeBench~\citep{jain2025livecodebench}, SWE-bench Verified~\citep{jimenez2024swebench},
DS-1000~\citep{lai2023ds1000}, Terminal-Bench 2.1, and claw-eval. These cover
text-to-SQL, competitive programming, software engineering, data-science coding,
terminal-based tasks, and agentic tool use. We use fixed evaluation sets, including
hard task subsets selected before adaptation; their sizes and construction procedures
are given in Appendix~\ref{app:tasks}.

We compare each task's fixed baseline harness with TTHE initialized from that
harness, using \texttt{deepseek-v4-flash} as the frozen solver, proposer, and judge.
Following Section~\ref{sec:setting}, TTHE adapts to each unlabeled batch; final
correctness feedback is used only to evaluate committed outputs. Metrics are task
success rates for the five binary benchmarks and mean graded scores for claw-eval.
Both methods are run three times per evaluation set. Figure~\ref{fig:main} shows
the means; individual results and mean $\pm$ sample standard deviation appear in
Appendix~\ref{app:repeated-results}.

\subsection{Main Results}
\label{subsec:main-results}
\label{sec:claw}
TTHE outperforms the fixed baseline on all six benchmarks in each of the three
reported runs (Figure~\ref{fig:main}; Appendix~\ref{app:repeated-results}). Mean
success rates increase by $8.0$--$36.0$ percentage points on the five binary
benchmarks; claw-eval gains $20.5$ points on its $0$--$100$ graded-score scale.

BIRD shows the largest gain, from $12.7\%$ to $48.7\%$ mean execution accuracy. Gains also occur on LiveCodeBench, SWE-bench Verified,
and DS-1000, and extend to terminal-based tasks and multi-service tool use:
Terminal-Bench 2.1 rises from $55.1\%$ to $67.8\%$, and claw-eval from $48.9\%$
to $69.4\%$. Across these settings, the same frozen backbone supports task execution,
program revision, and candidate selection. The complete procedure thus obtains
useful test-time harness updates without modifying model weights or introducing
a separate repair model.

\subsection{Ablation Studies}
\label{subsec:ablation}
We examine repeated sampling, search configuration, cross-batch initialization,
and backbone variation on BIRD. Unless stated otherwise, the ablations and
further analyses below use the main experiment's BIRD subset and
\texttt{deepseek-v4-flash}, reporting single-run accuracy separately from the
three-run main results. The reference run achieves $12.0\%$ with the fixed
baseline and $50.0\%$ with TTHE.

\paragraph{Repeated sampling without evolution.}
\label{subsec:repeated-sampling}
We run the fixed baseline ten times per question and vote over execution-result
groups rather than SQL strings. This matches TTHE's nominal $1+GR=10$ harness
executions for $G=3,R=3$: one initial execution and up to nine candidate executions.
Total computation also depends on harness-specific token and tool use, as well as
TTHE's proposal and judging overhead. Voting raises accuracy from $12.0\%$
to $16.0\%$, versus TTHE's $50.0\%$ (Table~\ref{tab:repeated-sampling}). The
contrast is between obtaining more outputs from an unchanged procedure and adapting
the procedure that produces and checks them. The tested voting strategy does not
substitute for the complete harness-evolution process.

\begin{table}[t]
\centering\small
\setlength{\tabcolsep}{5pt}
\renewcommand{\arraystretch}{1.15}
\setlength{\belowcaptionskip}{\abovecaptionskip}
\setlength{\abovecaptionskip}{0pt}
\caption{TTHE versus repeated sampling with a fixed harness.}
\label{tab:repeated-sampling}
\begin{tabular}{@{}lr@{}}
\toprule
Method & Accuracy (\%) \\
\midrule
Baseline (one execution) & 12.0 \\
Baseline (10 executions + vote) & 16.0 \\
TTHE ($G=3,R=3$) & \textbf{50.0} \\
\bottomrule
\end{tabular}
\end{table}

\paragraph{Search budget.}
Search breadth and depth interact (Table~\ref{tab:gr_ablation}). Increasing
$G$ from one to three does not improve accuracy at $R=1$ ($46.0\%$ versus
$44.0\%$), but improves it at $R=3$ ($40.0\%$ versus $50.0\%$).
Moreover, $G=3,R=1$ and $G=1,R=3$ allocate the same nominal number of proposals
per batch but achieve $44.0\%$ and $40.0\%$, respectively. These observations
suggest that search organization matters beyond proposal count, with $G=3,R=3$
the best tested setting. The branch configurations combine different search widths
with branch-specific proposer roles.

\begin{table}[t]
\centering\small
\setlength{\tabcolsep}{10pt}
\renewcommand{\arraystretch}{1.15}
\setlength{\belowcaptionskip}{\abovecaptionskip}
\setlength{\abovecaptionskip}{0pt}
\caption{Effects of search breadth and depth.}
\label{tab:gr_ablation}
\begin{tabular}{@{}lcc@{}}
\toprule
Accuracy (\%) & $G=1$ & $G=3$ \\
\midrule
$R=1$ & $46.0$ & $44.0$ \\
$R=3$ & $40.0$ & $\mathbf{50.0}$ \\
\bottomrule
\end{tabular}
\end{table}

\begin{figure}[t]
\centering
\includegraphics[width=0.9\linewidth]{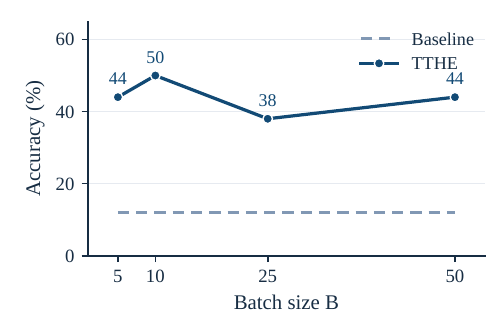}
\vspace{-0.3cm}
\caption{Effect of batch size on harness adaptation.}
\vspace{-0.4cm}
\label{fig:batch}
\end{figure}

\paragraph{Batch size.}
With $G=3,R=3$, accuracy peaks at $B=10$ ($50.0\%$) among the tested sizes,
versus $44.0\%$, $38.0\%$, and $44.0\%$ at $B=5$, $25$, and $50$
(Figure~\ref{fig:batch}). Batch size controls adaptation granularity: larger
batches provide more execution evidence, but one committed program must serve more
inputs and there are fewer update opportunities over the stream. The non-monotonic
results show that supplying more evidence per update is not uniformly beneficial
under this batch-level commitment rule.

\paragraph{Cross-batch accumulation.}
Resetting to the baseline before each batch achieves $44.0\%$ accuracy,
compared with $12.0\%$ for the fixed baseline. Carrying the selected harness
forward further improves overall accuracy to $50.0\%$
(Table~\ref{tab:accum}). These results indicate that current-batch
adaptation yields substantial gains, while retaining evolved program
structure provides a useful initialization for subsequent adaptation.

\begin{table}[t]
\centering\small
\setlength{\tabcolsep}{3.5pt}
\renewcommand{\arraystretch}{1.15}
\setlength{\belowcaptionskip}{\abovecaptionskip}
\setlength{\abovecaptionskip}{0pt}
\caption{Effect of cross-batch harness accumulation.}
\label{tab:accum}
\begin{tabular}{@{}lcccccc@{}}
\toprule
Accuracy (\%) & B1 & B2 & B3 & B4 & B5 & Overall \\
\midrule
Accumulate & 30 & 70 & 40 & 60 & 50 & \textbf{50.0} \\
Reset each batch & 30 & 40 & 40 & 50 & 60 & 44.0 \\
\bottomrule
\end{tabular}
\end{table}

\paragraph{Backbone variation.}
We repeat evolution on BIRD with MiMo\,V2.5 and Kimi\,K2.5, using the same backbone
across solver, proposer, and judge within each run ($B=10,G=3,R=3$).
Both improve over their own fixed-harness baselines: MiMo\,V2.5 increases from
$32.0\%$ to $52.0\%$, and Kimi\,K2.5 from $28.0\%$ to $48.0\%$
(Table~\ref{tab:crossmodel}). These within-model gains extend the evidence
beyond the main backbone. 

\begin{table}[t]
\centering\small
\setlength{\tabcolsep}{4pt}
\renewcommand{\arraystretch}{1.15}
\setlength{\belowcaptionskip}{\abovecaptionskip}
\setlength{\abovecaptionskip}{0pt}
\caption{Harness evolution with different backbones.}
\label{tab:crossmodel}
\begin{tabular}{@{}lccc@{}}
\toprule
Backbone & Base (\%) & TTHE (\%) & Gain \\
\midrule
DeepSeek V4 Flash & 12.0 & 50.0 & $+38.0$ \\
MiMo V2.5 & 32.0 & 52.0 & $+20.0$ \\
Kimi K2.5 & 28.0 & 48.0 & $+20.0$ \\
\bottomrule
\end{tabular}
\end{table}

\subsection{Further Analysis}
\label{subsec:further-analysis}
\paragraph{Cross-model harness transfer.}
\label{subsec:crossmodel-transfer}
Executing DeepSeek's selected harness on the corresponding BIRD batch, without
further evolution, improves MiMo\,V2.5 from $32.0\%$ to $56.0\%$ and GPT-5.5
from $36.0\%$ to $68.0\%$ (Table~\ref{tab:harness-transfer}). Backbone variation
tests whether other models can use the evolution procedure; this experiment instead
tests whether they can use its resulting programs. The gains show that a
batch-specific harness can benefit execution models other than the one that evolved
it.

\begin{table}[t]
\centering\small
\setlength{\tabcolsep}{4pt}
\renewcommand{\arraystretch}{1.15}
\setlength{\belowcaptionskip}{\abovecaptionskip}
\setlength{\abovecaptionskip}{0pt}
\caption{Cross-model transfer of evolved harnesses.}
\label{tab:harness-transfer}
\begin{tabular}{@{}lrrr@{}}
\toprule
Target model & \shortstack{Baseline\\(\%)} & \shortstack{Transferred\\(\%)} & Gain \\
\midrule
MiMo V2.5 & 32.0 & \textbf{56.0} & $+24.0$ \\
GPT-5.5 & 36.0 & \textbf{68.0} & $+32.0$ \\
\bottomrule
\end{tabular}
\end{table}

\paragraph{Candidate availability and selection.}
For the BIRD reference run, we compare the committed harnesses with two post-hoc
batchwise oracles. Each oracle selects the eligible harness with the most correct
outputs on each batch, using gold correctness for diagnosis only; neither mixes
different harnesses' outputs within a batch. The oracle over the final candidates
available to the judge reaches $64.0\%$ ($32/50$), compared with the committed
$50.0\%$ ($25/50$). The resulting $14$-point selection regret is a net deficit of
seven solved tasks. Trace audits include selecting a $3/10$ candidate over an
available $4/10$, and a $6/10$ candidate over a $7/10$. The audited errors include
executable SQL that returns extra columns or an inappropriate granularity,
illustrating the gap between execution success and task correctness.

The oracle over all evaluated candidates, including earlier rounds,
reaches $70.0\%$, six points above the final-pool oracle. Both oracles
use the candidate pools of the realized run. The fourteen-point gap
between the judge and the final-pool oracle exceeds the six-point
gain from admitting earlier candidates, highlighting selection among
already generated programs as a major source of lost performance.

\paragraph{What strategies emerge through evolution?}
Inspection of the evolved harnesses reveals concrete strategies for acquiring
evidence, directing repairs, and coordinating tool use. On BIRD, an evolved
harness probes database values before SQL generation, grounds filter values
in their stored casing, and infers the required output shape from the question.
Structural violations, execution errors, and output-shape mismatches are
returned as targeted feedback for subsequent attempts
(Figure~\ref{fig:evolved-harness}). Generation is therefore coupled with
explicit checks and failure-specific revision, rather than repeated without
diagnosing the preceding attempt. On SWE-bench, an evolved harness adopts a reproduce-first workflow with
root-cause localization, post-edit verification, and execution of existing
tests, alongside an increased step budget. It also adds a recovery branch
triggered when the initial rollout returns an empty patch. This branch
supplies concrete starter commands and a reproduction template, turning
the absence of a usable artifact into an explicit condition for changing
the execution strategy (Figure~\ref{fig:evolved-swe}). The harness thus
addresses not only how to verify a patch, but also how to resume execution
when the agent fails to produce one. On LiveCodeBench, an evolved policy
uses a low-cost path by default and escalates to additional verification
and repair after a public-test failure. In claw-eval, the evolved instructions address how information and actions
are linked across services. They specify single-line tool payloads, service
chaining through exact identifiers rather than display names, and a
pre-delivery completeness check requiring citations of retrieved record
identifiers. They also require reported actions to be supported by executed
tool calls. Across these examples, evolution changes
what evidence the agent obtains before acting, which observations trigger
revision, and how execution proceeds after failure. Execution experience
is incorporated into the rules governing task solving, rather than used
only to correct an individual answer.

\section{Limitations}
\label{sec:limitations}
TTHE demonstrates the potential of harness self-evolution without final
task-correctness supervision. However, the feedback guiding evolution
remains limited by the informativeness of execution traces and partial
checks. Successful execution or agreement among candidates can coexist
with semantic errors, making it difficult to distinguish effective
program improvements from changes that merely satisfy available proxies.
Our diagnostics show that useful candidates can already exist in the
search pool yet remain unselected. This highlights a central challenge:
an agent must draw on its own experience both to identify promising
changes and to assess whether those changes constitute genuine progress.
Developing principled mechanisms for extracting more informative feedback,
guiding program exploration, and selecting beneficial updates while
preserving useful behavior remains an important direction for advancing
reliable unsupervised agent self-evolution.
\section{Conclusion}
TTHE turns test-time execution experience into updates to an agent's executable
harness, without changing model weights or accessing final correctness feedback
during adaptation. Across six execution-grounded benchmarks, it improves over
fixed baseline harnesses on the evaluated task sets and produces inspectable
grounding, verification, and recovery policies. Transfer experiments further
show that batch-specific evolved harnesses can benefit other execution models
on the corresponding tasks. Our conditional analysis characterizes the balance
between corrective gains and unintended regressions for trace-triggered edits,
while search diagnostics show that generating useful candidates does not ensure
that they are selected. Together, these findings support executable harness
adaptation as a practical path to test-time self-evolution, with reliable
selection under incomplete evidence remaining a central challenge.

\clearpage
\bibliographystyle{plainnat}
\bibliography{refs}

@inproceedings{sun2020ttt,
  title={Test-time training with self-supervision for generalization under distribution shifts},
  author={Sun, Yu and Wang, Xiaolong and Liu, Zhuang and Miller, John and Efros, Alexei and Hardt, Moritz},
  booktitle={International Conference on Machine Learning, {ICML}},
  year={2020},
}

@inproceedings{wang2021tent,
  author       = {Dequan Wang and
                  Evan Shelhamer and
                  Shaoteng Liu and
                  Bruno A. Olshausen and
                  Trevor Darrell},
  title        = {Tent: Fully Test-Time Adaptation by Entropy Minimization},
  booktitle    = {International Conference on Learning Representations, {ICLR}},
  year         = {2021},
}

@article{wei2022cot,
  title={Chain-of-thought prompting elicits reasoning in large language models},
  author={Wei, Jason and Wang, Xuezhi and Schuurmans, Dale and Bosma, Maarten and Xia, Fei and Chi, Ed and Le, Quoc V and Zhou, Denny and others},
  journal={Advances in neural information processing systems},
  volume={35},
  pages={24824--24837},
  year={2022}
}

@inproceedings{wang2023selfconsistency,
  author       = {Xuezhi Wang and
                  Jason Wei and
                  Dale Schuurmans and
                  Quoc V. Le and
                  Ed H. Chi and
                  Sharan Narang and
                  Aakanksha Chowdhery and
                  Denny Zhou},
  title        = {Self-Consistency Improves Chain of Thought Reasoning in Language Models},
  booktitle    = {International Conference on Learning Representations, {ICLR}},
  year         = {2023},
}

@inproceedings{yao2023react,
  author       = {Shunyu Yao and
                  Jeffrey Zhao and
                  Dian Yu and
                  Nan Du and
                  Izhak Shafran and
                  Karthik R. Narasimhan and
                  Yuan Cao},
  title        = {ReAct: Synergizing Reasoning and Acting in Language Models},
  booktitle    = {International Conference on Learning Representations, {ICLR}},
  year         = {2023},
}

@inproceedings{chen2024selfdebug,
  author       = {Xinyun Chen and
                  Maxwell Lin and
                  Nathanael Sch{\"{a}}rli and
                  Denny Zhou},
  title        = {Teaching Large Language Models to Self-Debug},
  booktitle    = {International Conference on Learning Representations, {ICLR}},
  year         = {2024},
}

@inproceedings{shinn2023reflexion,
  author       = {Noah Shinn and
                  Federico Cassano and
                  Ashwin Gopinath and
                  Karthik Narasimhan and
                  Shunyu Yao},
  title        = {Reflexion: language agents with verbal reinforcement learning},
  booktitle    = {Advances in Neural Information Processing Systems, {NeurIPS}},
  year         = {2023},
}

@inproceedings{hu2025adas,
  author       = {Shengran Hu and
                  Cong Lu and
                  Jeff Clune},
  title        = {Automated Design of Agentic Systems},
  booktitle    = {International Conference on Learning Representations, {ICLR}},
  year         = {2025},
}

@inproceedings{zhang2025aflow,
  author       = {Jiayi Zhang and
                  Jinyu Xiang and
                  Zhaoyang Yu and
                  Fengwei Teng and
                  Xionghui Chen and
                  Jiaqi Chen and
                  Mingchen Zhuge and
                  Xin Cheng and
                  Sirui Hong and
                  Jinlin Wang and
                  Bingnan Zheng and
                  Bang Liu and
                  Yuyu Luo and
                  Chenglin Wu},
  title        = {AFlow: Automating Agentic Workflow Generation},
  booktitle    = {International Conference on Learning Representations, {ICLR}},
  year         = {2025},
}

@article{khattab2024dspy,
  title={{DSPy}: Compiling declarative language model calls into self-improving pipelines},
  author={Khattab, Omar and Singhvi, Arnav and Maheshwari, Paridhi and Zhang, Zhiyuan and Santhanam, Keshav and Vardhamanan, Sri and Haq, Saiful and Sharma, Ashutosh and Joshi, Thomas T and Moazam, Hanna and others},
  journal={arXiv preprint arXiv:2310.03714},
  year={2023}
}

@inproceedings{fernando2024promptbreeder,
  author       = {Chrisantha Fernando and
                  Dylan Banarse and
                  Henryk Michalewski and
                  Simon Osindero and
                  Tim Rockt{\"{a}}schel},
  title        = {Promptbreeder: Self-Referential Self-Improvement via Prompt Evolution},
  booktitle    = {International Conference on Machine Learning, {ICML}},
  year         = {2024},
}

@inproceedings{zhou2023ape,
  author       = {Yongchao Zhou and
                  Andrei Ioan Muresanu and
                  Ziwen Han and
                  Keiran Paster and
                  Silviu Pitis and
                  Harris Chan and
                  Jimmy Ba},
  title        = {Large Language Models are Human-Level Prompt Engineers},
  booktitle    = {International Conference on Learning Representations, {ICLR}},
  year         = {2023},
}

@article{zelikman2023stop,
  title={Self-taught optimizer ({STOP}): Recursively self-improving code generation},
  author={Zelikman, Eric and Lorch, Eliana and Mackey, Lester and Kalai, Adam Tauman},
  journal={arXiv preprint arXiv:2310.02304},
  year={2023}
}

@article{zhang2025dgm,
  title={Darwin godel machine: Open-ended evolution of self-improving agents},
  author={Zhang, Jenny and Hu, Shengran and Lu, Cong and Lange, Robert and Clune, Jeff},
  journal={arXiv preprint arXiv:2505.22954},
  year={2025}
}

@article{zhang2026selfharness,
  title={Self-Harness: Harnesses That Improve Themselves},
  author={Zhang, Hangfan and Zhang, Shao and Li, Kangcong and Zhang, Chen and Chen, Yang and Zhang, Yiqun and Bai, Lei and Hu, Shuyue},
  journal={arXiv preprint arXiv:2606.09498},
  year={2026}
}

@inproceedings{zelikman2022star,
  author       = {Eric Zelikman and
                  Yuhuai Wu and
                  Jesse Mu and
                  Noah D. Goodman},
  title        = {{STaR}: Bootstrapping Reasoning With Reasoning},
  booktitle    = {Advances in neural information processing systems, {NeurIPS}},
  year         = {2022},
}

@inproceedings{huang2023selfimprove,
  author       = {Jiaxin Huang and
                  Shixiang Gu and
                  Le Hou and
                  Yuexin Wu and
                  Xuezhi Wang and
                  Hongkun Yu and
                  Jiawei Han},
  title        = {Large Language Models Can Self-Improve},
  booktitle    = {Proceedings of the 2023 Conference on Empirical Methods in Natural
                  Language Processing, {EMNLP}},
  year         = {2023},
}

@article{he2026evotest,
  title={{EvoTest}: Evolutionary test-time learning for self-improving agentic systems},
  author={He, Yufei and Liu, Juncheng and Liu, Yue and Li, Yibo and Cao, Tri and Hu, Zhiyuan and Xu, Xinxing and Hooi, Bryan},
  journal={arXiv preprint arXiv:2510.13220},
  year={2025}
}

@inproceedings{li2023bird,
  title={Can {LLM} already serve as a database interface? a big bench for large-scale database grounded text-to-{SQL}s},
  author={Li, Jinyang and Hui, Binyuan and Qu, Ge and Yang, Jiaxi and Li, Binhua and Li, Bowen and Wang, Bailin and Qin, Bowen and Geng, Ruiying and Huo, Nan and others},
  booktitle={Advances in Neural Information Processing Systems, {NeurIPS}},
  year={2023}
}

@inproceedings{pourreza2023dinsql,
  title={{DIN-SQL}: Decomposed in-context learning of text-to-{SQL} with self-correction},
  author={Pourreza, Mohammadreza and Rafiei, Davood},
  booktitle={Advances in Neural Information Processing Systems, {NeurIPS}},
  year={2023}
}

@article{gao2024dailsql,
  title={Text-to-{SQL} empowered by large language models: A benchmark evaluation},
  author={Gao, Dawei and Wang, Haibin and Li, Yaliang and Sun, Xiuyu and Qian, Yichen and Ding, Bolin and Zhou, Jingren},
  journal={arXiv preprint arXiv:2308.15363},
  year={2023}
}

@inproceedings{jain2025livecodebench,
  author       = {Naman Jain and
                  King Han and
                  Alex Gu and
                  Wen{-}Ding Li and
                  Fanjia Yan and
                  Tianjun Zhang and
                  Sida Wang and
                  Armando Solar{-}Lezama and
                  Koushik Sen and
                  Ion Stoica},
  title        = {{LiveCodeBench}: Holistic and Contamination Free Evaluation of Large
                  Language Models for Code},
  booktitle={International Conference on Learning Representations, {ICLR}},
  year         = {2025},
}

@article{chen2023codet,
  title={{CodeT}: Code generation with generated tests},
  author={Chen, Bei and Zhang, Fengji and Nguyen, Anh and Zan, Daoguang and Lin, Zeqi and Lou, Jian-Guang and Chen, Weizhu},
  journal={arXiv preprint arXiv:2207.10397},
  year={2022}
}

@article{li2022alphacode,
  title={Competition-level code generation with {AlphaCode}},
  author={Li, Yujia and Choi, David and Chung, Junyoung and Kushman, Nate and Schrittwieser, Julian and Leblond, R{\'e}mi and Eccles, Tom and Keeling, James and Gimeno, Felix and Dal Lago, Agustin and others},
  journal={Science},
  year={2022},
}

@inproceedings{jimenez2024swebench,
  title={{SWE-bench}: Can language models resolve real-world {GitHub} issues?},
  author={Jimenez, Carlos E and Yang, John and Wettig, Alexander and Yao, Shunyu and Pei, Kexin and Press, Ofir and Narasimhan, Karthik},
  booktitle={International Conference on Learning Representations, {ICLR}},
  year={2024}
}

@inproceedings{yang2024sweagent,
  title={{SWE-agent}: Agent-computer interfaces enable automated software engineering},
  author={Yang, John and Jimenez, Carlos and Wettig, Alexander and Lieret, Kilian and Yao, Shunyu and Narasimhan, Karthik and Press, Ofir},
  booktitle={Advances in Neural Information Processing Systems, {NeurIPS}},
  year={2024}
}

@inproceedings{lai2023ds1000,
  title={{DS-1000}: A natural and reliable benchmark for data science code generation},
  author={Lai, Yuhang and Li, Chengxi and Wang, Yiming and Zhang, Tianyi and Zhong, Ruiqi and Zettlemoyer, Luke and Yih, Wen-tau and Fried, Daniel and Wang, Sida and Yu, Tao},
  booktitle={International Conference on Machine Learning, {ICML}},
  year={2023},
}

@article{lee2026metaharness,
  title={Meta-harness: End-to-end optimization of model harnesses},
  author={Lee, Yoonho and Nair, Roshen and Zhang, Qizheng and Lee, Kangwook and Khattab, Omar and Finn, Chelsea},
  journal={arXiv preprint arXiv:2603.28052},
  year={2026}
}

@article{moss2026,
  title={MOSS: Self-Evolution through Source-Level Rewriting in Autonomous Agent Systems},
  author={Cai, Qianshu and Zhang, Yonggang and Jia, Xianzhang and Xue, Wei and Song, Jun and Tian, Xinmei and Guo, Yike},
  journal={arXiv preprint arXiv:2605.22794},
  year={2026}
}

@inproceedings{yang2024opro,
  title={Large language models as optimizers},
  author={Yang, Chengrun and Wang, Xuezhi and Lu, Yifeng and Liu, Hanxiao and Le, Quoc V and Zhou, Denny and Chen, Xinyun},
  booktitle={International Conference on Learning Representations, {ICLR}},
  year={2024}
}

@inproceedings{guo2024evoprompt,
  title={Connecting large language models with evolutionary algorithms yields powerful prompt optimizers},
  author={Guo, Qingyan and Wang, Rui and Guo, Junliang and Li, Bei and Song, Kaitao and Tan, Xu and Liu, Guoqing and Bian, Jiang and Yang, Yujiu},
  booktitle={International Conference on Learning Representations, {ICLR}},
  year={2024}
}

@article{romeraparedes2024funsearch,
  title={Mathematical discoveries from program search with large language models},
  author={Romera-Paredes, Bernardino and Barekatain, Mohammadamin and Novikov, Alexander and Balog, Matej and Kumar, M Pawan and Dupont, Emilien and Ruiz, Francisco JR and Ellenberg, Jordan S and Wang, Pengming and Fawzi, Omar and others},
  journal={Nature},
  year={2024},
}

@inproceedings{madaan2023selfrefine,
  title={{Self-Refine}: Iterative refinement with self-feedback},
  author={Madaan, Aman and Tandon, Niket and Gupta, Prakhar and Hallinan, Skyler and Gao, Luyu and Wiegreffe, Sarah and Alon, Uri and Dziri, Nouha and Prabhumoye, Shrimai and Yang, Yiming and others},
  booktitle={Advances in Neural Information Processing Systems, {NeurIPS}},
  year={2023}
}

@inproceedings{zheng2023judging,
  author       = {Lianmin Zheng and
                  Wei{-}Lin Chiang and
                  Ying Sheng and
                  Siyuan Zhuang and
                  Zhanghao Wu and
                  Yonghao Zhuang and
                  Zi Lin and
                  Zhuohan Li and
                  Dacheng Li and
                  Eric P. Xing and
                  Hao Zhang and
                  Joseph E. Gonzalez and
                  Ion Stoica},
  title        = {Judging {LLM}-as-a-Judge with {MT-Bench} and Chatbot Arena},
  booktitle    = {Advances in Neural Information Processing Systems, {NeurIPS}},
  year         = {2023},
}

@article{wang2024voyager,
  author       = {Guanzhi Wang and
                  Yuqi Xie and
                  Yunfan Jiang and
                  Ajay Mandlekar and
                  Chaowei Xiao and
                  Yuke Zhu and
                  Linxi Fan and
                  Anima Anandkumar},
  title        = {Voyager: An Open-Ended Embodied Agent with Large Language Models},
  journal      = {Trans. Mach. Learn. Res.},
  year         = {2024},
}

@inproceedings{lightman2024verify,
  author       = {Hunter Lightman and
                  Vineet Kosaraju and
                  Yuri Burda and
                  Harrison Edwards and
                  Bowen Baker and
                  Teddy Lee and
                  Jan Leike and
                  John Schulman and
                  Ilya Sutskever and
                  Karl Cobbe},
  title        = {Let's Verify Step by Step},
  booktitle    = {International Conference on Learning Representations, {ICLR}},
  year         = {2024},
}

@inproceedings{zhang2022memo,
  author       = {Marvin Zhang and
                  Sergey Levine and
                  Chelsea Finn},
  title        = {{MEMO:} Test Time Robustness via Adaptation and Augmentation},
  booktitle    = {Advances in Neural Information Processing Systems, {NeurIPS}},
  year         = {2022},
}

@inproceedings{agrawal2026gepa,
  title={Gepa: Reflective prompt evolution can outperform reinforcement learning},
  author={Agrawal, Lakshya A and Tan, Shangyin and Soylu, Dilara and Ziems, Noah and Khare, Rishi and Opsahl-Ong, Krista and Singhvi, Arnav and Shandilya, Herumb and Ryan, Michael J and Jiang, Meng and others},
  booktitle={International Conference on Learning Representations},
  volume={2026},
  pages={8479--8565},
  year={2026}
}

@article{chen2026harnessx,
  title={Harnessx: A composable, adaptive, and evolvable agent harness foundry},
  author={Chen, Tingyang and Lu, Shuo and Zhao, Kang and Meng, Weicheng and Teng, Hanlin and Li, Tianhao and Li, Chao and Liu, Xule and Liang, Jian and Zhang, Zhizhong and others},
  journal={arXiv preprint arXiv:2606.14249},
  year={2026}
}

@article{luo2026hase,
  title={Harness-Aware Self-Evolving: Co-Evolving Model Weights, Harness, and Task Solutions},
  author={Luo, Haochen and Huang, Yi and Luo, Sichun and Liu, Fengyuan and Li, Lei and Hu, Zefa and Feng, Junlan and Liu, Qi},
  journal={arXiv preprint arXiv:2607.03935},
  year={2026}
}

@article{yang2026onerecipe,
  title={One Recipe, Many Harnesses: What Self-Evolution Encodes Across Languages and Models},
  author={Yang, Siqi and Yang, Qianlan and Wang, Yu-Xiong and Pujar, Saurabh and Hirzel, Martin},
  journal={arXiv preprint arXiv:2608.10178},
  year={2026}
}

@article{cai2026reliableharness,
  title={Safe Harness Self-Evolution: A Theoretical Analysis of Feasibility and Limits},
  author={Cai, Qianshu and Zhang, Yonggang and Nie, Jun and Ran, Maohao and Zheng, Huajiang and Song, Jun and Tian, Xinmei and Guo, Yike and Xue, Wei},
  journal={arXiv preprint arXiv:2609.08175},
  year={2026}
}

@article{zweiger2025seal,
  title={Self-adapting language models},
  author={Zweiger, Adam and Pari, Jyo and Guo, Han and Kim, Yoon and Agrawal, Pulkit},
  journal={Advances in Neural Information Processing Systems},
  volume={38},
  pages={74084--74115},
  year={2026}
}

@article{zhai2025agentevolver,
  title={Agentevolver: Towards efficient self-evolving agent system},
  author={Zhai, Yunpeng and Tao, Shuchang and Chen, Cheng and Zou, Anni and Chen, Ziqian and Fu, Qingxu and Mai, Shinji and Yu, Li and Deng, Jiaji and Cao, Zouying and others},
  journal={arXiv preprint arXiv:2511.10395},
  year={2025}
}

@article{zuo2025ttrl,
  title={Ttrl: Test-time reinforcement learning},
  author={Zuo, Yuxin and Zhang, Kaiyan and Sheng, Li and Qu, Shang and Cui, Ganqu and Zhu, Xuekai and Li, Haozhan and Long, Xinwei and Hua, Ermo and Qi, Biqing and others},
  journal={Advances in Neural Information Processing Systems},
  volume={38},
  pages={131459--131483},
  year={2026}
}

@article{hubotter2025ttcrl,
  title={Learning on the job: Test-time curricula for targeted reinforcement learning},
  author={H{\"u}botter, Jonas and Diaz-Bone, Leander and Hakimi, Ido and Krause, Andreas and Hardt, Moritz},
  journal={arXiv preprint arXiv:2510.04786},
  year={2025}
}

@article{yuksekgonul2026tttdiscover,
  title={Learning to discover at test time},
  author={Yuksekgonul, Mert and Koceja, Daniel and Li, Xinhao and Bianchi, Federico and McCaleb, Jed and Wang, Xiaolong and Kautz, Jan and Choi, Yejin and Zou, James and Guestrin, Carlos and others},
  journal={arXiv preprint arXiv:2601.16175},
  year={2026}
}

@inproceedings{suzgun2025dynamiccheatsheet,
  title={Dynamic cheatsheet: Test-time learning with adaptive memory},
  author={Suzgun, Mirac and Yuksekgonul, Mert and Bianchi, Federico and Jurafsky, Dan and Zou, James},
  booktitle={Proceedings of the 19th Conference of the European Chapter of the Association for Computational Linguistics (Volume 1: Long Papers)},
  pages={7080--7106},
  year={2026}
}

@inproceedings{zhang2026ace,
  title={Agentic context engineering: Evolving contexts for self-improving language models},
  author={Zhang, Qizheng and Hu, Changran and Upasani, Shubhangi and Ma, Boyuan and Hong, Fenglu and Kamanuru, Vamsidhar and Rainton, Jay and Wu, Chen and Ji, Mengmeng and Li, Hanchen and others},
  booktitle={International Conference on Learning Representations},
  volume={2026},
  pages={86069--86100},
  year={2026}
}

\clearpage
\onecolumn
\appendix
\setcounter{secnumdepth}{2}  

\section{Task Construction}
\label{app:tasks}
We use fixed evaluation sets that include selected hard task subsets. The four
query and code subsets were fixed before running TTHE using the criteria below;
no TTHE outcome was used to select or filter instances. Reference artifacts used
for subset construction were not passed to the adaptation loop.

\begin{table}[!htbp]
\centering\small
\setlength{\belowcaptionskip}{\abovecaptionskip}
\setlength{\abovecaptionskip}{0pt}
\caption{Evaluated task counts. The main results repeat each method three times
on its evaluation set.}
\label{tab:task-sets}
\begin{tabular}{@{}lrl@{}}
\toprule
Benchmark & Tasks per run & Metric \\
\midrule
BIRD & 50 & Execution accuracy \\
LiveCodeBench & 60 & Task success rate \\
SWE-bench Verified & 40 & Task success rate \\
DS-1000 & 50 & Task success rate \\
Terminal-Bench 2.1 & 89 & Task success rate \\
claw-eval & 30 & Mean graded task score \\
\bottomrule
\end{tabular}
\end{table}

\paragraph{BIRD.}
The 50-question hard subset is drawn from BIRD Mini-Dev by auditing questions
the baseline repeatedly missed and retaining unambiguous, substantively hard
cases. It spans multiple databases. Accuracy uses result-set equality between the
executed prediction and the reference query. Selection is conditioned on baseline
failures; the scores describe this recovery-oriented subset, not full-benchmark
performance or a controlled ranking of different backbones.

\paragraph{LiveCodeBench.}
The 60-problem hard subset uses the contamination-controlled release window
reported in the experiment, with problems published after the solver's training
cutoff. Task success requires passing all hidden tests; exposed examples remain
available as partial checks during adaptation.

\paragraph{SWE-bench Verified.}
The 40-instance hard subset is selected by gold-patch complexity, using multi-file
or large changes. The baseline is the off-the-shelf mini-swe-agent scaffold,
which TTHE subsequently evolves~\citep{yang2024sweagent}. Reference patches used
for subset construction are not given to the adaptation loop.

\paragraph{DS-1000.}
The 50-problem hard subset is selected by reference-solution length. Reference
solutions and final evaluation tests are excluded from adaptation; reported
success requires passing the hidden task tests.

\paragraph{Terminal-Bench 2.1.}
The main experiment evaluates 89 tasks from Terminal-Bench 2.1. The benchmark
repository is \url{https://github.com/harbor-framework/terminal-bench-2-1}.

\paragraph{claw-eval.}
The 30-task headroom subset consists of tasks served by local deterministic mock
services and selected using baseline performance. Appendix~\ref{app:claw} describes
the subset construction, the OpenClaw harness, graded scoring, and the information
boundary. The baseline and TTHE each have three complete-run results, listed in
Appendix~\ref{app:repeated-results}.

\subsection{Model Access}
\label{app:model-apis}
Table~\ref{tab:model-apis} lists the API base URLs for the models used in our experiments.

\begin{table}[!htbp]
\centering\small
\renewcommand{\arraystretch}{1.15}
\setlength{\belowcaptionskip}{\abovecaptionskip}
\setlength{\abovecaptionskip}{0pt}
\caption{Model API endpoints.}
\label{tab:model-apis}
\begin{tabular}{@{}ll@{}}
\toprule
Model & API base URL \\
\midrule
DeepSeek V4 Flash & \url{https://api.deepseek.com} \\
MiMo V2.5 & \url{https://api.xiaomimimo.com/v1} \\
Kimi K2.5 & \url{https://api.moonshot.cn/v1} \\
GPT-5.5 & \url{https://api.openlux.ai/v1} \\
\bottomrule
\end{tabular}
\end{table}

\subsection{Dataset Licenses}
\label{app:dataset-licenses}
Table~\ref{tab:dataset-licenses} summarizes the licensing information provided by
upstream dataset cards and benchmark repositories. Third-party task content remains
subject to its original license or terms of use.

\begin{table}[!htbp]
\centering\small
\setlength{\tabcolsep}{5pt}
\renewcommand{\arraystretch}{1.2}
\setlength{\belowcaptionskip}{\abovecaptionskip}
\setlength{\abovecaptionskip}{0pt}
\caption{Benchmark licenses and source terms.}
\label{tab:dataset-licenses}
\begin{tabular}{@{}p{0.23\linewidth}p{0.48\linewidth}p{0.21\linewidth}@{}}
\toprule
Benchmark & License / terms & Source \\
\midrule
BIRD Mini-Dev & CC BY-SA 4.0 & \href{https://huggingface.co/datasets/birdsql/bird_mini_dev}{Dataset card} \\
\addlinespace[2pt]
LiveCodeBench & Creative Commons (\texttt{cc} in the dataset card; variant unspecified) & \href{https://huggingface.co/datasets/livecodebench/code_generation_lite}{Dataset card} \\
\addlinespace[2pt]
SWE-bench Verified & Source-repository terms (task content); MIT (benchmark code) & \href{https://huggingface.co/datasets/SWE-bench/SWE-bench_Verified}{Dataset card}; \href{https://github.com/SWE-bench/SWE-bench/blob/main/LICENSE}{code license} \\
\addlinespace[2pt]
DS-1000 & CC BY-SA 4.0 & \href{https://huggingface.co/datasets/xlangai/DS-1000}{Dataset card} \\
\addlinespace[2pt]
Terminal-Bench 2.1 & Apache 2.0 & \href{https://github.com/harbor-framework/terminal-bench-2-1/blob/main/LICENSE}{Repository license} \\
\addlinespace[2pt]
claw-eval & MIT & \href{https://huggingface.co/datasets/claw-eval/Claw-Eval}{Dataset card} \\
\bottomrule
\end{tabular}
\end{table}

\section{Discovered Harness Policies}
The evolved programs are retained as source code rather than summarized only by aggregate scores.
Recurring discovered policies include:
\begin{itemize}
    \item \textbf{Text-to-SQL}: value grounding through database probes, date-format repair, explicit
    output-shape checks, and removal of accidental limits on list queries.
    \item \textbf{LiveCodeBench}: multiple candidate programs, execution-based ranking on public tests,
    targeted repair from expected-versus-observed outputs, and fresh-start recovery after repeated failure.
    \item \textbf{SWE-bench}: reproduce-first debugging, root-cause localization before editing,
    post-edit verification, and an anti-paralysis failsafe for issues that would otherwise return an empty
    patch (Figure~\ref{fig:evolved-swe}).
    \item \textbf{DS-1000}: insertion-only output contracts, deterministic self-checks, and guards against
    redefining supplied variables.
    \item \textbf{claw-eval}: single-line tool payloads so write actions are recorded, per-record
    identifier citation with a pre-delivery completeness check, exact-identifier service chaining, and a
    prohibition on reporting unexecuted actions (see the claw-eval section below).
\end{itemize}
These policies are examples of program-level adaptation: they modify the control procedure used across
instances rather than appending a one-off answer correction.

\begin{figure}[!htbp]
\centering
\scriptsize
\begin{varwidth}{\linewidth}
\begin{verbatim}
RULES = ("Hint is authoritative; use exact DB casing (SQLite"
    " '=' is case-sensitive); return EXACTLY the asked"
    " column(s); superlative -> ORDER BY .. LIMIT 1 with a"
    " tiebreaker; no ROUND() unless asked; NULLIF safe div.")

class EvolvedSQLHarness(SQLHarness):
  def solve(self, question):
    # (1) ground filter values vs the DB (exact casing)
    grounding = self.ground_values(question)
    # (2) infer expected output shape (#cols, 1 vs many)
    shape = self.explain_shape(question)
    # (3) generate -> validate -> run -> self-debug (<=6x)
    err = ""
    for _ in range(6):
      sql = extract_sql(self.llm(
          self.prompt(question, grounding, shape, err),
          system=RULES))
      if (msg := self.validate_structure(sql, question)):
          err = msg; continue        # e.g. SELECT *
      res = self.execute(sql)
      if not res["ok"]:
          err = "SQL error: " + res["error"]; continue
      if not res["rows"]:
          err = "0 rows -- recheck WHERE casing"; continue
      if (msg := self.check_columns(question, res)):
          err = msg; continue        # wrong / extra columns
      return self.fix_shape(question, sql, res)
    return sql
\end{verbatim}
\end{varwidth}
\caption{Source of an evolved Text-to-SQL harness (excerpt, lightly abridged), discovered from label-free
execution traces alone. It grounds filter values against the database for exact casing, infers the expected
output shape from the question, and wraps generation in an execute-and-self-debug loop that returns each
failure a targeted error message---structural check, execution error, empty result, wrong column
count---before a final shape fix. The baseline is a standard ReAct harness (before any test-time evolution).}
\label{fig:evolved-harness}
\end{figure}

\begin{figure}[!htbp]
\centering
\scriptsize
\begin{varwidth}{\linewidth}
\begin{verbatim}
# Evolved SWE-bench harness (excerpt, abridged). Its header records
# what the proposer synthesized from 15 traces of earlier candidates:
#  - reproduce-first works: 4/5 issues yield a patch once the
#    agent starts running commands;
#  - one issue yields EMPTY patches across ALL harnesses: the
#    agent is "paralyzed" and emits no commands at all;
#  - "previous attempt failed" messaging deepens the freeze;
#  - the first 3-5 commands decide success.
# Synthesized policy: reproduce-first + anti-paralysis failsafe.

_SYSTEM = """
You are a focused engineer fixing a bug from a PR description.
## Mandatory workflow (IN ORDER, never skip)
  1. REPRODUCE FIRST: run /tmp/repro.py BEFORE reading source;
     you MUST see the wrong behaviour.
  2. TRACE with grep/sed; never cat whole files.
  3. DECLARE root cause BEFORE editing, citing the buggy lines.
  4. FIX MINIMALLY: 1-10 lines in ONE source file (never tests/).
  5. VERIFY: re-run /tmp/repro.py; the bug output MUST be gone.
  6. TEST: run pytest once.   7. SUBMIT the diff.
## Momentum is critical
  If you go 5 steps without running a command you are STUCK: run
  ANY command to break the logjam.
"""
# A separate anti-paralysis phase gives a 4-command starter
# (ls -> check Python -> find models -> write repro) and a
# copy-paste repro template, and never mentions prior failure.
\end{verbatim}
\end{varwidth}
\caption{An evolved SWE-bench harness (excerpt, abridged), starting from the stock mini-swe-agent and
discovered from label-free execution traces alone. The proposer's own comments record the failure mode it
found---some issues drive the agent into ``empty-patch paralysis''---and the reproduce-first workflow and
anti-paralysis failsafe it synthesized in response. None of this was supplied.}
\label{fig:evolved-swe}
\end{figure}

\FloatBarrier
\section{The claw-eval Agentic Tool-Use Domain}
\label{app:claw}

\subsection{Task, metric, and protocol}
claw-eval tasks are multi-service office workflows: the solver reads mail, searches a knowledge base,
looks up contacts, inspects a CRM or helpdesk, schedules meetings, and updates tickets through tool calls.
The score is a per-task graded rubric in $[0,1]$, rather than binary execution accuracy.
Each rubric combines completion, robustness, and communication criteria
under a safety factor; a held-out model grader assesses the committed behavior.

The $30$ tasks come from $139$ agentic tasks: we retain the $112$ served entirely by
local deterministic mock services, then select those on which the frozen baseline
scores below $0.60$. The solver is \texttt{deepseek-v4-flash} inside the OpenClaw
agent, and the proposer edits its system prompt and skills as source code.
The run processes the tasks in six batches, evolving one general harness from the
baseline. The grader is disabled during in-loop observation, so no task score or
rubric target enters the proposer or judge; gold grading is applied only to the
committed harness for measurement. The baseline and TTHE scores, $48.9\%$ and $69.4\%$,
are each averaged over three runs (Appendix~\ref{app:repeated-results}). The tasks are ordered by ascending baseline difficulty, so this evaluation does not
isolate the effect of cross-batch accumulation.

\subsection{Discovered policies}
From label-free traces alone the
accumulated harness adds four general rules to the baseline: single-line tool payloads for reliable
action capture; citation of every retrieved record identifier under a completeness checklist before
delivery; chaining services by exact identifier (email or ID) rather than display name; and grounding
every reported action in an executed tool call.

\section{Related Work}

\paragraph{Prompt and workflow optimization.}
Research on optimizing the system around a language model spans instructions,
workflows, and executable programs. APE and OPRO search for
instructions~\citep{zhou2023ape,yang2024opro}, while Promptbreeder and EvoPrompt
use evolutionary prompt optimization~\citep{fernando2024promptbreeder,guo2024evoprompt}.
DSPy optimizes declarative language-model pipelines against a
metric~\citep{khattab2024dspy}; ADAS and AFlow extend the search to agent designs
and code-represented workflows~\citep{hu2025adas,zhang2025aflow}. FunSearch
illustrates the broader potential of executable program
search~\citep{romeraparedes2024funsearch}. More recently, GEPA uses natural-language
reflection on reasoning and tool-use trajectories to propose prompt updates and
combine complementary improvements from a Pareto frontier~\citep{agrawal2026gepa}.
Thus, interpreting execution traces is already an important optimization mechanism;
TTHE studies its use for revising and selecting an executable harness on the
current test batch without final task-correctness feedback.

\paragraph{Executable harness evolution.}
A related line modifies agent source code or accumulates executable components.
STOP and the Darwin G\"odel Machine revise agent
scaffolds~\citep{zelikman2023stop,zhang2025dgm}, Voyager builds a reusable skill
library~\citep{wang2024voyager}, and MOSS studies source-level rewriting from
failure trajectories~\citep{moss2026}. Among harness-level methods, Self-Harness
validates edits through verifier-grounded evaluation and regression tests with a
fixed model~\citep{zhang2026selfharness}; Meta-Harness gives its proposer access
to historical code, scores, and traces from a search set before held-out
evaluation~\citep{lee2026metaharness}. HarnessX provides composable harness
primitives and a trace-driven multi-agent evolution engine, with both
harness-only adaptation and a harness--model co-evolution loop that reuses
trajectories for model training~\citep{chen2026harnessx}. HASE unifies task-solution
generation and harness editing in an agentic reinforcement-learning process,
co-evolving model weights, harness components, and
solutions~\citep{luo2026hase}. TTHE shares the executable-program optimization
target but keeps the solver, proposers, and judge frozen, and excludes final
correctness feedback from both current-batch program revision and commitment.

\paragraph{Transfer and reliable adoption.}
Beyond aggregate performance, recent work examines what evolved harnesses retain
and when their updates can be trusted. One Recipe, Many Harnesses analyzes
artifacts produced by a common evolution recipe across languages and models,
distinguishing transferable strategies from ecosystem-specific
adaptations~\citep{yang2026onerecipe}. A theory of reliable harness self-evolution
studies adoption conditions that balance gains on failed tasks against preservation
of successful behavior, together with the limits imposed by verification and
evaluation costs~\citep{cai2026reliableharness}. TTHE's transfer experiments instead
execute batch-specific evolved programs with other backbones. Its conditional
analysis concerns a class of trace-triggered edits under a fixed execution coupling.

\paragraph{Self-directed model adaptation.}
Self-improvement can also change the model itself. STaR bootstraps reasoning by
fine-tuning on generated rationales that lead to correct
answers~\citep{zelikman2022star}; related work fine-tunes on high-confidence
solutions generated from unlabeled questions~\citep{huang2023selfimprove}.
SEAL learns to generate its own adaptation data and update directives, using
post-update performance to train the self-editing policy~\citep{zweiger2025seal}.
AgentEvolver combines self-generated tasks, experience-guided exploration, and
trajectory-level credit assignment to improve agent reinforcement
learning~\citep{zhai2025agentevolver}. These approaches use experience to improve
parameterized capabilities or the process that trains them. In TTHE, experience
instead changes the executable procedure through which an unchanged model
acquires context, uses tools, verifies results, and recovers from failures.

\paragraph{Test-time adaptation and reinforcement learning.}
TTT, Tent, and MEMO adapt model parameters or normalization statistics using
self-supervised, confidence-based, or augmentation-based objectives on unlabeled
test inputs~\citep{sun2020ttt,wang2021tent,zhang2022memo}. For language-model
reasoning, TTRL performs reinforcement learning directly on unlabeled test data,
estimating rewards from agreement with majority-voted model
outputs~\citep{zuo2025ttrl}. It therefore shares the absence of ground-truth
answers during adaptation, but uses the inferred rewards to update model weights.
TTC-RL constructs test-time curricula from an available training pool for targeted
reinforcement learning~\citep{hubotter2025ttcrl}, while TTT-Discover trains on
experience from a particular test problem, using continuous rewards to prioritize
exceptional solutions~\citep{yuksekgonul2026tttdiscover}. These studies establish
test-time learning as more than repeated inference. TTHE pursues a different
adaptation state: the model remains fixed, while execution evidence informs
program updates and candidate selection without being reduced to a scalar
training objective.

\paragraph{Test-time memory and context evolution.}
Parameter-free adaptation is also studied through persistent memory and evolving
contexts. Dynamic Cheatsheet maintains a self-curated memory of strategies,
code snippets, and reusable insights at inference time without explicit
ground-truth labels~\citep{suzgun2025dynamiccheatsheet}. ACE treats contexts as
evolving playbooks, using generation, reflection, and curation with incremental
updates; it supports both offline context optimization and online adaptation
from execution feedback~\citep{zhang2026ace}. EvoTest evolves prompts, memory,
hyperparameters, and executable state-extraction logic across interactive
episodes, using transcripts for updates and environment-score statistics for
candidate selection~\citep{he2026evotest}. These methods already demonstrate
experience accumulation beyond an isolated answer. TTHE focuses on revising the
task's control program and committing a single harness whose cached outcomes
are scored on the same batch, with final evaluation feedback excluded from
the adaptation loop.

\paragraph{Response refinement and execution feedback.}
Chain-of-thought, self-consistency, and ReAct organize reasoning and tool use
within an inference procedure~\citep{wei2022cot,wang2023selfconsistency,yao2023react}.
Self-debugging and Self-Refine revise candidate
answers~\citep{chen2024selfdebug,madaan2023selfrefine}, and Reflexion retains
textual experience for later attempts~\citep{shinn2023reflexion}. Domain-specific
systems incorporate these ideas into fixed procedures: DIN-SQL and DAIL-SQL use
structured prompting and execution-related processing for
text-to-SQL~\citep{pourreza2023dinsql,gao2024dailsql}, while code-generation
systems use tests to select candidate solutions~\citep{chen2023codet,li2022alphacode}.
Learned verifiers and LLM judges provide further evaluation
signals~\citep{lightman2024verify,zheng2023judging}. TTHE can express verification
and repair within a harness, but adapts the program that specifies when and how
these operations occur, rather than only revising a particular response.

\section{Cross-Database Transfer and Continued Adaptation}
\label{app:database-transfer}

TTHE targets adaptation to the current unlabeled batch. As a complementary
evaluation, we examine whether an evolved harness transfers to databases
excluded from its adaptation stage, and whether continued adaptation yields
further gains on the target batches.

\paragraph{Setup and protocol.}
We partition the baseline-screened \texttt{genuine\_hard50} subset of BIRD
Mini-Dev~\citep{li2023bird} by database, preserving the original question
order within each split. The ID split contains 30 questions from
\texttt{card\_games}, \texttt{formula\_1}, and \texttt{toxicology};
the OOD split contains 20 questions from \texttt{california\_schools},
\texttt{superhero}, \texttt{debit\_card\_specializing}, and
\texttt{student\_club}. Here, OOD denotes database-disjoint transfer within
the same screened benchmark subset.

Starting from the bare text-to-SQL harness, TTHE evolves on the ID split.
The final ID harness is frozen before exposure to any OOD questions or
evaluation scores. We evaluate this fixed program on both splits without
further evolution. Separately, we initialize TTHE with the frozen ID
harness and continue adaptation on two OOD batches, scoring each batch
with its corresponding committed harness.

All runs use \texttt{deepseek/deepseek-v4-flash} with temperature $0$,
reasoning enabled, and two workers. TTHE uses batch size $B=10$,
$G=3$ branches, and at most $R=3$ rounds. Solver and proposer timeouts
are $180$ and $2700$ seconds, respectively.


\begin{table}[!htbp]
\centering
\small
\renewcommand{\arraystretch}{1.15}
\setlength{\belowcaptionskip}{\abovecaptionskip}
\setlength{\abovecaptionskip}{0pt}
\caption{Cross-database transfer and continued harness adaptation.}
\label{tab:database-transfer}
\begin{tabular}{@{}llrr@{}}
\toprule
Evaluation split & Harness setting & Correct / Total & Accuracy (\%) \\
\midrule
ID  & Bare baseline                       & 5 / 30  & 16.7 \\
ID  & Frozen ID harness                   & 10 / 30 & 33.3 \\
\midrule
OOD & Bare baseline                       & 2 / 20  & 10.0 \\
OOD & Frozen ID harness                   & 10 / 20 & 50.0 \\
OOD & Continued TTHE from the ID harness  & 13 / 20 & \textbf{65.0} \\
\bottomrule
\end{tabular}
\end{table}

\paragraph{Results and interpretation.}
The frozen ID harness achieves $10/30$ on the ID adaptation set,
compared with $5/30$ for the bare baseline
(Table~\ref{tab:database-transfer}). More importantly, it achieves
$10/20$ on the held-out databases without further evolution, versus
$2/20$ for the baseline. This improvement provides evidence that the
evolved task-solving procedure remains useful beyond the databases
that supplied its adaptation traces.

Continuing TTHE on the OOD split raises accuracy to $13/20$, with
$8/10$ and $5/10$ on the two successive batches. The frozen-transfer
result measures reuse without target-side harness evolution, whereas
the continued-adaptation result measures the additional benefit of
updating the procedure using target-batch execution evidence.
Together, these results support both cross-database reuse of an
evolved harness and its capacity for further test-time adaptation.
\section{Proofs and Scope of the Conditional Analysis}
\label{app:theory}

\subsection{Probability Space and the Locality Condition}
\paragraph{Fixed construction history and fresh executions.}
Fix the batch, its hidden targets, and a realization $z$ of the label-free construction history and
random choices defining $H,H'$ and the trigger rule $\mathsf d$. Program configurations and inherited
state are held fixed. Execution randomness is drawn anew from each program's prescribed law; it is not
a replay or a conditional reweighting of the rollouts that produced $z$. The uniform index $I$ averages
over this same batch, so no i.i.d. assumption about a test stream is needed.

For task $i$, let $E_i$ denote its prescribed initial environment and let $P_H^z(\cdot\mid x_i,E_i)$
be the joint law of the output and trace from the fixed program $H$, including any randomized checks
recorded in that trace. Define $P_{H'}^z$ analogously. A specified coupling $\kappa$ must satisfy
\[
\begin{aligned}
 \mathcal L_\kappa(O,\tau\mid I=i)&=P_H^z(\cdot\mid x_i,E_i),\\
 \mathcal L_\kappa(O',\tau'\mid I=i)&=P_{H'}^z(\cdot\mid x_i,E_i).
\end{aligned}
\]
The two executions start from the same prescribed task state, not from a state modified by the other
execution. All probabilities below are for this fixed history and coupling. In particular,
$D=\mathsf d(x_I,\tau)$ has its actual joint law with $O$; preserving the output marginal alone would
not suffice. A child check, when used, is likewise a fixed function of $x_I$ and $\tau'$, with any
check randomness included in its record. Hidden targets enter only the correctness indicators.
Complete outcomes include artifacts and final states that affect scoring, so $O=O'$ implies $C=C'$.

\paragraph{Coupling and a prefix-preserving repair.}
The error rates $e,e'$ and their difference depend only on the prescribed marginals, whereas $r,u$ and
$\varepsilon$ can depend on $\kappa$. For example, let $H'=H$, set $D=0$, and suppose the program
returns either of two outcomes with equal probability. Sharing its random draw gives $\varepsilon=0$;
independent draws give $\varepsilon=1/2$, although $\Delta_X=0$ in both comparisons. Thus
$\varepsilon$ is not an intrinsic distance between source programs.

For the gated structure illustrated in the main text, couple the child and parent by sharing the
unchanged parent execution, including its trace and random draws. The child returns its outcome when
$D=0$ and invokes repair with fresh randomness when $D=1$. This coupling has the required marginals
only if the child actually implements that prefix-preserving behavior. The check must preserve all
score-relevant untriggered outcomes. Probes and repairs must not change the target, interfere with other
instances, or consume resources that alter untriggered behavior. Adding an \texttt{if} statement alone
does not establish these properties. No independence between the two executions is required.

\paragraph{Cached outcomes and adaptive selection.}
Construction may use the same batch, but the realized rollouts in $z$ are distinct from the fresh
execution laws above. Conditioning on a program chosen using those rollouts need not preserve their
original sampling law. One can also decompose correctness transitions in fixed cached outcomes by
averaging over batch indices, but that is an empirical identity for those outcomes, not a guarantee
for fresh executions or the adaptive selection process. The results here neither rank candidates from
unlabeled evidence nor ensure that the proposer or judge returns a pair satisfying the stated conditions.

\subsection{Proof of Proposition~\ref{prop:balance}}
\begin{proof}
Split the correctness gain according to whether the trigger fires:
\begin{equation}
 \Delta_X(H,H')=\mathbb E[C'-C]
 =\mathbb E[(C'-C)\mathbb I[D=1]]
 +\mathbb E[(C'-C)\mathbb I[D=0]].
 \label{eq:proofsplit}
\end{equation}
On the triggered region, only incorrect-to-correct and correct-to-incorrect transitions contribute:
\begin{align}
 \mathbb E[(C'-C)\mathbb I[D=1]]
 &=\Pr(C=0,D=1,C'=1)-\Pr(C=1,D=1,C'=0)\\
 &=m\pi r-m(1-\pi)u.
 \label{eq:triggerbalance}
\end{align}
The products follow from the chain rule, without independence. If $m=0$, both triggered event masses
vanish; otherwise the conditioning events partition $D=1$. The convention for zero-probability events
therefore preserves the identity.

Let $\Gamma_0=\mathbb E[(C'-C)\mathbb I[D=0]]$. Since identical outcomes have the same correctness,
\begin{equation}
 |\Gamma_0|\leq\mathbb E[|C'-C|\mathbb I[D=0]]
 \leq\Pr(D=0,O'\ne O)=\varepsilon.
 \label{eq:offtriggerbound}
\end{equation}
Thus $\Delta_X(H,H')=m[\pi r-(1-\pi)u]+\Gamma_0$, which proves Eq.~\ref{eq:editbound}.
If $m\pi r>m(1-\pi)u+\varepsilon$, its lower bound is positive. With $\varepsilon=0$, the equality
makes $m\pi r>m(1-\pi)u$ necessary and sufficient for strict improvement.
\end{proof}

\paragraph{Equivalent error-detection-rate parameterization.}
Define $d=\Pr(D=1\mid C=0)$ and $f=\Pr(D=1\mid C=1)$, with the same zero-event convention.
Here $d$ is error detection and $f$ is false triggering. The joint event masses satisfy
\[
 m\pi=ed,\qquad m(1-\pi)=(1-e)f.
\]
Consequently, Proposition~\ref{prop:balance} is equivalently
\begin{equation}
 \left|e'-\big[e(1-dr)+(1-e)fu\big]\right|\leq\varepsilon.
 \label{eq:detectionbalance}
\end{equation}
In particular, $edr>(1-e)fu+\varepsilon$ suffices for improvement. This is the same probability balance,
expressed in terms of error detection and false triggering rather than trigger mass and precision.

\paragraph{Exact locality and sharpness of the uniform coefficient.}
For any $e,d,f,r,u\in[0,1]$, the following nonnegative masses sum to one:
\begin{center}
\begin{tabular}{cccc}
\toprule
$C$ & $D$ & $C'$ & Probability mass\\
\midrule
$0$ & $1$ & $1$ & $edr$\\
$0$ & $1$ & $0$ & $ed(1-r)$\\
$0$ & $0$ & $0$ & $e(1-d)$\\
$1$ & $1$ & $0$ & $(1-e)fu$\\
$1$ & $1$ & $1$ & $(1-e)f(1-u)$\\
$1$ & $0$ & $1$ & $(1-e)(1-f)$\\
\bottomrule
\end{tabular}
\end{center}
Choose identical outcomes on the two $D=0$ rows. Exact locality then holds, and the correctness gain is
$edr-(1-e)fu=m[\pi r-(1-\pi)u]$. Parameters attached to zero-mass conditioning events do not affect the
distribution; their actual conditional rates are interpreted by the zero-event convention. This
construction realizes the exact balance when $\varepsilon=0$.

The coefficient of the uniform symmetric $\varepsilon$-term cannot be reduced in general: take a
parent correct everywhere, set $D=0$ everywhere, and change the child to an incorrect outcome on a
mass $\varepsilon$. Then $\Delta_X=-\varepsilon$, attaining the lower bound. Starting from an
everywhere-incorrect parent and repairing a mass $\varepsilon$ off-trigger attains the upper bound.
This is a uniform coefficient claim, not a claim that the symmetric interval is pointwise sharp for
every feasible parameter tuple; restrictions on the off-trigger error masses can give tighter bounds.

\subsection{Proof of Theorem~\ref{thm:soundrepair}}
\begin{proof}
Sound rejection gives $\Pr(D=1,C=1)=0$, so the regression event in Eq.~\ref{eq:triggerbalance} vanishes.
Exact locality gives $\Gamma_0=0$. Therefore
\[
 \Delta_X(H,H')=\Pr(C=0,D=1,C'=1)=\Pr(D=1,C'=1)=mr.
\]
When $m>0$, soundness gives $\pi=1$; when $m=0$, the expression is zero under the zero-event convention.
This proves preservation without assuming a positive repair success rate or independent executions.

If $D'$ is sound for rejection on the child, $C'=1$ implies $D'=0$ almost surely. Hence
\[
 \{D=1,C'=1\}\subseteq\{D=1,D'=0\}
\]
up to a null set. Taking probabilities yields $0\leq\Delta_X(H,H')\leq v$.

For sharpness, fix $0\leq v\leq m\leq1$ and construct three disjoint events of masses $1-m$, $m-v$, and
$v$. On the first, set $D=D'=0$, use identical outcomes, and set $C=C'=1$. On the second, set
$D=D'=1$ and $C=C'=0$. On the third, set $D=1$, $D'=0$, and $C=0$. For the lower endpoint set
$C'=0$ on this last event; for the upper endpoint set $C'=1$. Both constructions have the same
$(D,D')$ distribution, the same baseline error $e=m$, exact locality, and sound rejection for both
checks. Their correctness gains are zero and $v$, respectively. Checks can realize this by rejecting a
malformed outcome while accepting both a correct outcome and a well-formed but semantically wrong one.
The indistinguishability concerns these checks' outcomes, not all semantic information in a complete
execution trace.
\end{proof}

\subsection{Interpretation and Application Boundaries}
\paragraph{Failure soundness is one-sided.}
A sound rejection flags a \emph{final outcome} that cannot satisfy the task. Acceptance need not imply
correctness. A runtime failure is sound rejection only when it invalidates the submitted outcome under
the specified environment. Visible tests qualify only when they are authoritative necessary conditions;
noisy, stale, unrelated, or flaky tests do not automatically qualify. A correct SQL query may return zero
rows, and an intermediate exception may be recovered from before a correct outcome is submitted. Such
signals require potentially imperfect precision $\pi$, rather than being classified as sound by default.

\paragraph{Repair precision and locality.}
The mass $m\pi$ identifies incorrectly behaving executions targeted by an edit; $m(1-\pi)$ identifies
correct ones exposed to modification. A high trigger rate alone says nothing about their balance.
Without a repair-quality assumption, $r\geq0$ and $u\leq1$ give
\[
 e'\leq e+m(1-\pi)+\varepsilon=e+(1-e)f+\varepsilon.
\]
Soundness removes falsely triggered correct outcomes, and exact locality removes possible off-trigger
harm. Neither condition asserts $r>0$. Likewise, clearing a failure flag does not measure $r$, because
correct and incorrect outputs can both pass an incomplete check.

\paragraph{Relation to the search procedure.}
TTHE can express local repair branches as well as global prompt and grounding edits. The preservation
result applies to transitions satisfying the stated soundness and locality conditions; extending it to
a sequence requires these conditions at each committed transition. The complete search procedure is
evaluated empirically.

\paragraph{Binary versus graded outcomes.}
The analysis uses binary correctness, as in exact SQL or hidden-test evaluation. For graded loss,
merely detecting an outcome that is not fully correct does not make replacing it safe: a partly
successful outcome can be made worse. An analogous preservation conclusion would require a stronger
condition, such as the trigger certifying maximal loss. It is not claimed for claw-eval's graded score,
nor does the correctness analysis account for additional latency, tokens, or execution costs.

\section{Full Algorithm}
\label{app:algorithm}

\begin{algorithm}[H]
\small
\caption{Test-Time Harness Evolution (one stream)}
\label{alg:tthe}
\begin{algorithmic}
\STATE \textbf{Input:} solver $M$; stream $x_{1:T}$; batch size $B$; branches $G$; rounds $R$; baseline $H_0$
\STATE $H \leftarrow H_0$ \hfill // ReAct baseline harness
\FOR{each unlabeled batch $X$ of size $B$ in $x_{1:T}$}
\STATE $(\mathcal{T}[H],\mathcal{Y}[H]) \leftarrow \textsc{Observe}(H, X)$ \hfill // traces and cached outputs
\STATE $(H_0^1,\dots,H_0^G) \leftarrow (H,\dots,H)$ \hfill // $G$ branches from $H$, sharing its cached trace $\mathcal{T}[H]$
\FOR{$r=1$ to $R$}
\STATE $\mathcal{A} \leftarrow \{H_{r-1}^g\}_{g=1}^{G}$ \hfill // current active branches
\STATE $\{\tilde{H}_r^g\}_{g=1}^{G} \leftarrow \textsc{Propose}(\mathcal{A}, \mathcal{T}[\mathcal{A}])$ \hfill // branch $g$ edits its own parent $H_{r-1}^g$
\STATE $H_r^g \leftarrow \tilde{H}_r^g$ if valid, else $H_{r-1}^g$ \hfill // invalid child falls back to parent
\STATE $(\mathcal{T}[H_r^g],\mathcal{Y}[H_r^g]) \leftarrow \textsc{Observe}(H_r^g,X)$ for each newly generated child
\ENDFOR
\STATE $\mathcal{F} \leftarrow \{H_R^g\}_{g=1}^{G}$ \hfill // final branches only
\STATE $H \leftarrow \textsc{Judge}(\mathcal{F}, \mathcal{T}[\mathcal{F}], X)$ \hfill // label-free selection
\STATE $\textsc{Score}(\mathcal{Y}[H], \mathrm{Gold}(X))$ \hfill // measurement only; no re-execution
\ENDFOR
\end{algorithmic}
\end{algorithm}
\section{Repeated Main Results}
\label{app:repeated-results}
Table~\ref{tab:main-repeats} reports three runs per method with mean $\pm$ sample
standard deviation. Raw results are solved/total counts for the binary benchmarks
and mean graded scores for claw-eval. Figure~\ref{fig:main} uses the run means,
with absolute gains computed before rounding. Ablations and selection diagnostics
use the single-run reference described in Section~\ref{subsec:ablation}.

\begin{table}[!htbp]
\centering\small
\setlength{\tabcolsep}{6pt}
\renewcommand{\arraystretch}{1.15}
\setlength{\belowcaptionskip}{\abovecaptionskip}
\setlength{\abovecaptionskip}{0pt}
\caption{Run-level main results and mean $\pm$ sample standard deviation.}
\label{tab:main-repeats}
\begin{tabular}{@{}llrrrr@{}}
\toprule
Benchmark & Method & Run 1 & Run 2 & Run 3 & Mean $\pm$ SD (\%) \\
\midrule
BIRD & Baseline & 6/50 & 6/50 & 7/50 & $12.67\pm1.15$ \\
BIRD & TTHE & 25/50 & 24/50 & 24/50 & $48.67\pm1.15$ \\
\addlinespace[2pt]
LiveCodeBench & Baseline & 18/60 & 16/60 & 19/60 & $29.44\pm2.55$ \\
LiveCodeBench & TTHE & 23/60 & 24/60 & 21/60 & $37.78\pm2.55$ \\
\addlinespace[2pt]
SWE-bench Verified & Baseline & 8/40 & 9/40 & 9/40 & $21.67\pm1.44$ \\
SWE-bench Verified & TTHE & 14/40 & 13/40 & 14/40 & $34.17\pm1.44$ \\
\addlinespace[2pt]
DS-1000 & Baseline & 19/50 & 16/50 & 17/50 & $34.67\pm3.06$ \\
DS-1000 & TTHE & 22/50 & 21/50 & 21/50 & $42.67\pm1.15$ \\
\addlinespace[2pt]
Terminal-Bench 2.1 & Baseline & 49/89 & 46/89 & 52/89 & $55.06\pm3.37$ \\
Terminal-Bench 2.1 & TTHE & 60/89 & 63/89 & 58/89 & $67.79\pm2.83$ \\
\addlinespace[2pt]
claw-eval & Baseline & 42.78\% & 51.18\% & 52.70\% & $48.89\pm5.34$ \\
claw-eval & TTHE & 69.76\% & 65.21\% & 73.29\% & $69.42\pm4.05$ \\
\bottomrule
\end{tabular}
\end{table}

\end{document}